\documentclass[12pt,a4paper]{article}
\usepackage{amsmath}
\usepackage{graphicx}
\usepackage{times}
\begin{document}
\begin{titlepage}
\title{Implications of the   $\rho(s)$ measurements by TOTEM at the LHC}
\author{ S.M. Troshin, N.E. Tyurin\\[1ex]
\small  \it SRC IHEP of NRC ``Kurchatov Institute''\\
\small  \it Protvino, 142281, Russian Federation}
\normalsize
\date{}
\maketitle

\begin{abstract}
Implications of the recent measurements of the parameter $\rho(s)$ (ratio of the real to imaginary parts of the forward scattering elastic amplitude) by TOTEM collaboration at $\sqrt{s}=13$ $TeV$ are discussed with emphasis on the rising energy dependence of the ratio of elastic to total cross--sections.
\end{abstract}
\end{titlepage}
\setcounter{page}{2}
\section*{Introduction}
It is well known that dynamics of elastic $pp$--scattering at the LHC energies (where one can presumably use an assumption on the vanishing dependence on the spin degrees of freedom) is described by one complex function $F(s,t)$ of the two Mandelstam variables $s$ and $t$. These variables have different physical meaning in the direct and annihilation reaction channels and are equally important. The data of many experiments show that the scattering amplitude dependence of the two variables cannot be factorized. Such factorization of the amplitude contradicts also to analyticity and unitarity \cite{gribov}.   Obviously, the models for the description of the scattering  dynamics should  provide an explicit knowledge of  the scattering amplitude $F(s,t)$ as a function of both  variables. 

Instead of that, it is an often practice to consider an energy dependence of the scattering amplitude  based on its $s$--dependency at one  fixed value of the transferred momentum only, namely, at the momentum transfer value $t=0$. This choice is usually argued by the fact that such a global characteristic of hadron interaction dynamics as the total cross--section is determined by the value (of the imaginary part) of the scattering amplitude in forward direction, i.e. at $-t=0$.  

In the impact parameter representation, where the scattering amplitude is written as Fourier--Bessel transform
\begin{equation}\label{rfst}
 F(s,t)=\frac{s}{\pi^2}\int_0^{\infty}bdb f(s,b)J_0(b\sqrt{-t}),
\end{equation}
such approach corresponds to modeling the result of integration of the function $f(s,b)$ over impact parameter $b$ but not the function $f(s,b)$ itself. It is evident that it is not an equivalent to construction of the dynamical model leading to the explicit forms of the functions $f(s,b)$ or $F(s,t)$. Due to this ambiguity (since different non-factorized functions  can provide the same resulting function being integrated over one of independent variables)\footnote{A particular illustrating example can be found in \cite{dremw}.} there is no much sense in fitting  data by such ``models'', aka analytic parameterizations, since the important dynamics is  overintegrated and cannot therefore be disentangled. It is not clear what confident conclusions  can be made after comparison of their predictions with  the experiments. Thus, using  them as the reference  points to generate statements on the interaction dynamics or deviation from an assumed one is doubtful  since those have no unambiguous  ground. This remark is valid for any kind of functions but is mostly relevant for the potentially sign changing ones, for example, the real part of the elastic  amplitude.

The above introductory comment emphasizes importance of the both $s$-- and $t$-- oriented direct studies in the elastic scattering measurements at the LHC energies. 
The recent results of the TOTEM Collaboration  on the measurements of the differential cross-section of elastic scattering at small values of $-t$
  and $\sqrt{s}=8$ TeV \cite {totem} have demonstrated a new  interesting feature  of the elastic amplitude associated with an observation of the deviation from a simple exponential dependence of $d\sigma/dt\sim \exp{at}$. In another words, the quantity 
  \[
 B=\frac {d}{dt} \ln \frac{d\sigma}{dt}
  \]
 is a nontrivial function of $t$ in the region where a simple exponential dependence of $ d\sigma/dt$ was usually adopted.
  This feature has been discussed and quantitatively described using the two--exponential parameterization in \cite{real}.
  
  Recently, the new data on  the parameter $\rho(s)$ have been published \cite{totro} and  demonstrated a significant energy dependence of this parameter relevant to  forward $pp$-- scattering. Namely,  $\rho(s)$ has revealed a decreasing energy dependence  from $\sqrt{s}=8$ $TeV$ to $\sqrt{s}=13$ $TeV$. It can  be in favor  of a significant crossing--odd contribution into the elastic scattering amplitude and can be considered as an experimental confirmation  of the Maximal Odderon concept \cite{nicol,khoze}. 
  
  However,  the $\rho$ remains to be positive as it should be in case of the crossing--even contribution dominance \cite{martwu} while the Maximal Odderon contribution predicts significant negative energy--independent asymptotic value $-0.2$ for the ratio $\rho(s)$ \cite{nicol}.
  As it was noted in \cite{totro} the  $\rho(s)$ measurements would indicate an expected slow down of the total cross-section increase at higher energies in the case when the crossing-even contribution dominance is assumed. This conclusion was based on the local derivative relations for the $\rho(s)$  obtained in \cite{ddr} (from the integral dispersion relations):
  \begin{equation}\label{ddr}
  \rho(s)\sim \frac{d \ln \sigma_{tot}(s)}{d\ln s}.
  \end{equation}
  
  In this note we concentrate  on further implications of the TOTEM result: the energy dependence of the parameter $\rho(s)$ and its interrelations with the two parts of the total cross-section --- integrated contribution of elastic scattering along with the cross-section of all the inelastic collisions and expected  slow down of the total cross--section increase.

\section{The ratio $\rho(s)$ at the LHC energies} 
We start  with a brief overview of the results obtained in \cite{real} where scenario for the real part restoration has been proposed. It is based on  derivation of the real part of the scattering amplitude from the corresponding imaginary part through the derivative dispersion relation. It was noted that it seems appropriate to use term analytization for such procedure. 

To construct an imaginary part of the scattering amplitude in the impact parameter representation it was proposed to use the rational form of unitarization and the input function $U$ was assumed to be a pure imaginary, $U\to iU$:
\begin{equation}\label{ssbr}
f(s,b)=\frac{U(s,b)}{1+U(s,b)}.
\end{equation}
It  gives a  pure imaginary scattering amplitude $f\to if$. The input function $U(s,b)$ was chosen to have a factorized form on the base of geometrical considerations:
\begin{equation}
U(s,b)=g(s)\omega(b)
\end{equation}
with $g(s)\sim s^\lambda$ to guarantee the $\ln ^2 s$ increase of the total cross--section, and account of correct analytical properties of scattering amplitude is provided through the function $\omega (b)$. It should be noted that the resulting amplitude $f(s,b)$ is not a factorized function of its variables as well as the amplitude $F(s,t)$.  Energy dependence of the slope parameter $B(s)$ is generated by the unitarity and it has an asymptotic dependence $B(s)\sim \ln ^2 s$ \cite{mpla17}. 

To restore the real part of the scattering amplitude the  derivative dispersion relations in the impact parameter representation have been applied and the following relation has been obtained:
\begin{equation}\label{real}
\mbox{Re} F(s,t)\sim H_{inel}(s,t),
\end{equation}
where $H_{inel}(s,t)$ is the inelastic overlap function, or  a contribution of all the intermediate inelastic channels into  unitarity \cite{vanhove}.  The relation Eq. (\ref{real}) is, of course, a model dependent one and  based on the local version of the dispersion relations valid in high energy limit (hopefully at the LHC energies) \cite{ddr,anis}, but the assumptions used are based on general principles of the quantum field theory such as unitarity 
and analyticity along with the existing experimental trends.

We use Eq. (\ref{real}) for discussion of the energy dependence of  $\rho(s)$  in view of the new TOTEM measurements of this parameter. Its straightforward consequence  is proportionality of the real part of the scattering amplitude at $-t=0$ and the integral cross-section of the inelastic interactions, i.e.
\begin{equation}\label{real1}
\mbox{Re} F(s,-t=0)\sim s\sigma_{inel}(s)
\end{equation}
since $H_{inel}(s,-t=0)\sim s\sigma_{inel}(s)$. 
 The resulting energy dependence of $\rho(s)$
 \begin{equation}\label{ro}
 \rho(s)\sim \frac{\sigma_{inel}(s)}{\sigma_{tot}(s)}
 \end{equation}
 obeys the Khuri-Kinoshita theorem \cite{kinosh}, i.e.  it  decreases like $1 /\ln s$ since 
$\sigma_{tot}(s)\sim \ln^2 s$ while $\sigma_{inel}(s)\sim \ln s$  at  $s\to\infty$ due to the unitarity saturation \cite{usect}.  It is also in agreement with the recent Martin-Wu result on  positivity of the $\rho(s)$ \cite{martwu}. Thus, everything seems to be quite consistent. 

Eq. (\ref{ro}) and the available data \cite{totel} allow one to recalculate the  value of $\rho(s)$ at $\sqrt{s}=8$ $TeV$ from  the known ratio of  the cross--sections 
\[
r_i(s)\equiv {\sigma_{inel}(s)}/{\sigma_{tot}(s)}
\]
 at the energy value $\sqrt{s}=8$ $TeV$ and the data at $\sqrt{s}=13$ $TeV$. Taking from the data the ratio $r_i(s)=0.74$ at $\sqrt{s}=8$ $TeV$  and
$r_i(s)=0.72$ at $\sqrt{s}=13$ $TeV$ we obtain that $\rho(s)=0.103$ at $\sqrt{s}=8$ $TeV$ if $\rho(s)=0.100$ at $\sqrt{s}=13$ $TeV$. This value is in agreement with the experimental data (within the error bars), cf.  left panel of  Fig. 18 from \cite{totro}.

The ratio $r_i(s)$ demonstrates a slow decrease when the energy increases from 8 to 13 $TeV$. Since  the inelastic cross--section provides a major contribution to $\sigma_{tot}(s)$ in this energy range and is expected at higher energies,
the total cross--section would slow  down its increase at higher energies  due to decreasing  $r_i(s)$ and despite of increasing 
\[
r_e(s)\equiv\sigma_{el}(s)/\sigma_{tot}(s)
\]
 (evidently, $r_e(s)+r_i(s)=1$).
The latter ratio increases with energy and tends to its asymptotic value unity at $s\to\infty$  which reflects  unitarity saturation. 
It would be helpful to obtain the data at the  energy of $\sqrt{s}=8$ $TeV$ with better precision. Currently, we can  state for certain that the data for $\rho(s)$ at the LHC energies exclude decreasing behavior of the ratio $r_e(s)$. It ultimately corresponds to the experimental trends demonstrating the increasing energy dependence of $r_e(s)$ \cite{totel}.
\section*{Conclusion}
The above considerations lead to the relation $\rho(s)\sim r_i(s)$, and numerical estimates give  an increase of the ratio $r_e(s)$ at the energy $\sqrt{s}=13$ $TeV$ compared to its value at $\sqrt{s}=8$ $TeV$.  This increasing energy dependence of $r_e(s)$ also observed experimentally \cite{totel}  would result in its turn in transitory  slow down of the growth of $\sigma_{tot}(s)$ at this energy range and just beyond the LHC energies due to major  contribution ($>50\%$) of $\sigma_{inel}(s)$ in this energy range, but asymptotically monotonic $\ln^2 s$  dependence of $\sigma_{tot}$ would be finally observed. 

To conclude, we note again  that decreasing behavior of $\rho{(s)}$ is interrelated with increasing ratio $r_e(s)$ of the elastic to total cross--sections.
\section*{Acknowledgements}
We are grateful to E.S. Martynov for the discussion on the remote negative asymptotic value of $\rho(s)$ in the presence of contribution from the Maximal Odderon. 


\small
\begin{thebibliography}{99}
\bibitem{gribov}
V.N. Gribov, Proc. of the 1960 Int. Conf. on High Energy Physics, Rochester (1960), p. 340.
\bibitem{dremw}
I.M. Dremin, S.N. White,  arXiv:1604.03469v2.
\bibitem{totem}
G. Antchev et al. (The TOTEM Collaboration),  Nucl. Phys. B. {\bf 899} (2015) 527. 
\bibitem{real}
S.M. Troshin, N.E. Tyurin, Mod. Phys. Lett. A, {\bf 32} (2017) 1750028.
\bibitem{totro}
G. Antchev et al. (The TOTEM Collaboration), CERN-EP-2017-335.
\bibitem{nicol}
E. Martynov, B. Nicolescu, Phys. Lett. B {\bf 778} (2018) 414.
\bibitem{khoze}
V.A. Khoze, A.D. Martin, M.G. Ryskin, Phys. Rev. D {\bf 97} (2018) 034019.
\bibitem{martwu}
A. Martin, T.T. Wu, Phys. Rev. D {\bf 97} (2018) 014011.
\bibitem{ddr}
J.B. Bronzan, G.L. Kane, U.P. Sukhatme, Phys. Lett. B {\bf 49} (1974) 272.
\bibitem{anis}
V.V. Anisovich, V.A. Nikonov, J. Nyiri, Int. J. Mod. Phys. A, {\bf 30} (2015) 1550188.
\bibitem{mpla17}
S.M. Troshin, N.E. Tyurin, Mod. Phys. Lett. A {\bf 32} (2017) 1750168.
\bibitem{vanhove} L. Van Hove, 
Nuovo Cim. {\bf 28} (1963) 301.
\bibitem{kinosh}
N.N. Khuri, T. Kinoshita, Phys. Rev. B {\bf 137} (1965) 720.
\bibitem{usect}
S.M. Troshin, N.E. Tyurin, Int. J. Mod. Phys. A {\bf 22} (2007) 4437. 
\bibitem{totel}
G. Antchev et al. (The TOTEM Collaboration), arXiv: 171206153v2.

\end{thebibliography}
\end{document}